\documentclass[fleqn,usenatbib]{mnras}

\usepackage{newtxtext,newtxmath}
\usepackage[T1]{fontenc}
\usepackage{subcaption}
\usepackage{adjustbox}
\usepackage{threeparttable}
\usepackage{comment}

\DeclareRobustCommand{\VAN}[3]{#2}
\let\VANthebibliography\thebibliography
\def\thebibliography{\DeclareRobustCommand{\VAN}[3]{##3}\VANthebibliography}

\usepackage{graphicx}	% Including figure files
\usepackage{amsmath}	% Advanced maths commands
\title[CL AGNs discovered through optical variability]{Confirming new changing--look AGNs discovered through optical variability using a random-forest based light curve classifier}
\author[E. López-Navas et al.]{
E. López-Navas,$^{1}$\thanks{E-mail: elena.lopez@postgrado.uv.cl}
M.L. Martínez-Aldama,$^{1,2}$
S. Bernal,$^{1}$
P. Sánchez-Sáez,$^{3,4}$
P. Arévalo,$^{1}$
\newauthor
Matthew J. Graham,$^{5}$
L. Hernández-García,$^{4,1}$
P. Lira,$^{2}$ and
P.A. Rojas Lobos$^{1}$.
\\
$^{1}$Instituto de Física y Astronomía, Facultad de Ciencias, Universidad de Valparaíso, Gran Bretaña 1111, Valparaíso, Chile\\
$^{2}$Departamento Astronomía, Universidad de Chile, Casilla 36D, Santiago, Chile\\
$^{3}$European Southern Observatory, Karl-Schwarzschild-Strasse 2, 85748 Garching bei München, Germany\\
$^{4}$Millennium Institute of Astrophysics (MAS), Nuncio Monseñor Sótero Sanz 100, Providencia, Santiago, Chile \\
$^{5}$California Institute of Technology, 1200 E. California Blvd, Pasadena, CA 91125, USA
}

\date{Accepted XXX. Received YYY; in original form ZZZ}

\pubyear{2022}

\begin{document}
\label{firstpage}
\pagerange{\pageref{firstpage}--\pageref{lastpage}}
\maketitle

\begin{abstract}
%Changing--Look (CL) Active Galactic Nuclei (AGNs), where the optical broad emission lines appear or disappear, most likely correspond to large changes in their accretion rate. Determining the frequency and duration of these events is therefore crucial to understand the evolution of the accretion flow around supermassive black holes but it is hindered by the extreme rarity of these events and the expensive spectroscopic confirmation. Methods to find CL candidates efficiently and reliably are therefore key to establish the properties of this population. We present a strategy to select new CL candidates starting from a spectroscopic type 2 AGNs sample and searching for current type 1 photometric variability. We use the publicly available Zwicky Transient Facility (ZTF) alert stream and the Automatic Learning for the Rapid Classification of Events (ALeRCE) light curve classifier to produce a candidate list with a highly automated algorithm. We estimate the success rate of the selection method with new spectroscopic observations of six candidates, finding the appearance of clear broad Balmer lines in four of them and tentative evidence of type changes in the remaining two, when compared to their archival spectra. The selection method therefore shows a promising success rate of $\geq66$ per cent. 
Determining the frequency and duration of  changing--look (CL) active galactic nuclei (AGNs) phenomena, where the optical broad emission lines appear or disappear,  is crucial to understand the evolution of the accretion flow around supermassive black holes. We present a strategy to select new CL candidates starting from a spectroscopic type 2 AGNs sample and searching for current type 1 photometric variability. We use the publicly available Zwicky Transient Facility (ZTF) alert stream and the Automatic Learning for the Rapid Classification of Events (ALeRCE) light curve classifier to produce a list of CL candidates with a highly automated algorithm, resulting in 60 candidates. Visual inspection reduced the sample to 30. We performed new spectroscopic observations of six candidates of our clean sample, without further refinement, finding the appearance of clear broad Balmer lines in four of them and tentative evidence of type changes in the remaining two, which suggests a promising success rate of $\geq66$ per cent for this CL selection method.

\end{abstract}


%%%%%%%%%%%%%%%%%%%%%%%%%%%%%%%%%%%%%%%%%%%%%%%%%%

%%%%%%%%%%%%%%%%% BODY OF PAPER %%%%%%%%%%%%%%%%%%

\section{Introduction}

One remarkable property of active galactic nuclei (AGNs) is their variability, which is seen in
a wide range of the electromagnetic spectrum. Of particular interest among AGNs with extreme
variability are the changing-look (CL) AGNs, which display an appearance or disappearance of their
optical broad emission lines (BELs) on time scales from months to years. Since the broad line
region (BLR) is photoionized by the UV/optical continuum from the accretion disc \citep[e.g. see the review by][]{2015netzer}
 dramatic variations in this waveband often occur simultaneously with the
appearance/disappearance of the BELs, with a time-lag that is consistent with expectations from reverberation mapping studies  \citep[i.e light-days to light-weeks,][]{trakhtenbrot2019}. %Catching an AGN during a change of state is extremely rare but it provides the opportunity to study accretion physics at its most critical regimes, i.e. advancing/receding accretion discs and the formation of the BLR clouds.
In the Unification Model \citep[UM, e.g.][]{antonucci1993}, AGNs can be classified as type 1 or type 2 depending on the orientation of the system. Type 1 refers to AGNs whose nuclei are visible, while in type 2 AGNs a dusty structure (i.e., the torus) is expected to obscure a direct view to the accretion disc and the BLR, resulting in the absence of the disc’s blue continuum and of the BELs in the spectrum. This configuration also hides the intrinsic variability of the disc emission, so the optical flux in classical type 2 AGNs should be constant \citep{sanchez2017}. However, the existence of CL AGNs challenges the UM, as the presence of BELs should be determined by the inclination of the source, which cannot change drastically in a few year’s timescales.\\
The origin of the CL phenomenon in the optical range is still unclear,  but nearly all observational tests have now disfavoured transient dust obscuration (as in CL AGNs in the X--ray range) or nuclear tidal disruption events (TDEs) as the source of most of the observed fading/brightening of the BELs and continuum emission \citep{lamassa2015,runnoe2016,macleod2016,macleod2019,hutsemekers2017,hutsemekers2019}. Interestingly, these sources could be suffering dramatic changes intrinsic to the accretion flow,  in a way similar to what we observe when X ray--binaries (XRB) go into outburst \citep[e.g.][]{homan2005}. In this scenario, the CL AGN would make a spectral transition to/from an extreme UV bright accretion disc to a hot inner flow, which would produce a change in the BELs as a consequence of the changing shape of the ionizing spectrum. %When the accretion rate increases or decreases, the luminosity of the accretion disc suffers a dramatic change, and the BELs appear or disappear. Thus the AGN goes from an unobscured type 2 to a type 1 or vice versa. 
Recently, several studies have found evidence that supports the accretion state change as the physical origin of the CL phenomenon, although the observed time scale for this phenomenon is much shorter than expected if it was analogous to XRB outbursts. On the contrary, the events occur on thermal time scales, which has been associated with cooling/heating fronts propagating through the disc \citep{Noda2018,parker2019,graham2020, sniegowska2020}. Thus, if type 2 CL AGNs change their type due to a change in the accretion state, they must belong to the unobscured, \textit{true type 2} AGN population \citep[see][and references therein]{2001tran}.
% it has been demonstrated that the CL standard accretion disc could develop very quickly ($<$a year), likely driven by disc instabilities (Parker et al., 2019).  %(10$^{4\--5}$ years) 

%%%%changing state - not changing look and how many

In recent years, $\sim$ 200 CL AGNs have been successfully identified with a variety of methods, starting with blind searches of BELs variability by comparing archival optical spectra from different epochs \citep{yang2018,2022green}. However, most of the CL AGNs have been confirmed via follow-up spectroscopy of candidates that had experienced great variations in any of their properties: in the optical flux \citep{yang2018,frederick2019,macleod2019}, in the color \citep{2021hon}, and/or in the optical and mid-IR photometric variability \citep{graham2020}. These studies have shown the CL phenomenon is extremely rare. In a blind search for CL AGNs, \citet{yang2018} found that just 19 out of 330795 (the 0.006 per cent) galaxies with repeated spectroscopy in the SDSS and/or LAMOST are CL AGNs. From an initial sample of 1.1 millions quasars, \citet{graham2020} identified 47451 candidates that met specific optical and mid-IR photometric variability requirements, and found that 111 of them had significant spectral changes (the 15 per cent of their final sample with second epoch spectra). Since the origin of this variability is consistent with a change of state of activity and is not necessarily associated with a change in the optical type but in the flux of the BELs, these sources are called "changing-state" quasars. From this and previous works that include variability criteria to find new CL AGNs \citep{yang2018, sanchez2021_anomaly}, it is clear that a diversity of phenomena can lead to extreme optical variations that are not associated with significant spectroscopic changes, and other observables are required to investigate further this behaviour.

Optimising the method to find new CL candidates will help to improve the statistics on the frequency and duration of this phenomenon. With the advent of real time, deep, large sky-coverage monitoring surveys as the Zwicky Transient Facility \citep[ZTF,][]{bellm2019} it is possible now, for the first time, to obtain a census of the rate of these changes \citep{sanchez2021_anomaly}. Here, we find new CL using the machine learning classifications provided by the Automatic Learning for the Rapid Classification of Events \citep[ALeRCE,][]{forster2021} broker, which allows us to select a sample of 60 candidates that transitioned from type 2 to type 1 AGNs. To test our selection technique, we performed spectroscopic follow-up for six candidates, finding clear evidence of the appearance of BELs for at least four of them. Throughout this work, we assume a standard cosmological model with H$_0$ = 70 km s$^{-1}$ Mpc$^{-1}$, $\Omega_{m}$ = 0.3, and $\Omega_{\Lambda}$ = 0.7.

\section{Selection of the sample}\label{sec:sample} 
Our selection strategy is based on variability considerations: if an AGN shows rapid and strong optical variations we could have a direct view of its accretion disc and also of the BLR-- if it exists. Therefore, current type 1 variability seen in a spectrally-classified type 2 AGN could mean that the AGN changed type since its spectrum was taken.

In this work we used the current variability-based, publicly available classifications provided by the  ALeRCE light curve classifier tool \citep[LCC,][]{sanchez2021}. The LCC uses a hierarchical imbalanced Random Forest classifier, fed with variability features computed from ZTF light curves and colors obtained from AllWISE and ZTF photometry, to classify each source with ZTF alerts into 15 subclasses, including variable stars, transient events, and three classes of AGNs (host-dominated or AGN, core-dominated or QSO, and Blazar). The LCC is able to separate QSO, Blazar and AGN from other stochastic sources with a 99 percent success rate. For the classification of AGN-like sources, the most relevant features are the ZTF/ALLWISE colors, features related to the amplitude of the variability at different timescales, and features related to the timescale of the variability.

%, the New Catalog of Type 1 AGNs (Oh et al. 2015) and the Roma-BZCAT Multi-Frequency Catalog of Blazars (Massaro et al. 2015)

In particular, the ZTF produces alerts of all sufficiently variable objects (5 $\sigma$ variation in the template-subtracted images). Only 10 per cent of known type 1 AGNs with $r<21$ mag show alerts in the ZTF because the typical variability of these objects is too small to reach this threshold. Therefore, if former type 2 AGNs with current type 1 variability behave in the same way as the rest of the type 1 population, we only expect to detect 10 per cent of the CLs in the sample with these data and method.

Our initial sample consists of 42027 AGNs that were classified as type 2 in the Million Quasars Catalog \citep[MILLIQUAS,Version 7.1, N and K types, ][]{milliquas} or the Veron Catalog of Quasars \& AGN \citep[VERONCAT, 13th Edition, S2 and Q2 types, ][]{veron-cetty2010}. From these, we discarded sources that have been classified as Seyfert I, Low Ionization Nuclear Emission Region (LINER) or Blazar in any other study according to the SIMBAD Astronomical Database, and those without public "GALAXY" or "QSO" spectra in the SDSS database, nor subclassified as BROADLINE, which led to 22380 sources.%Thus, this method targets obscured type 1 and true type 2 AGNs indistinctly. The classification of CL events into these two initial categories will be addressed with a larger sample in future work. }
 %\textcolor{red}{( with "GALAXY" or "QSO" specclass in the SDSS databasewithin 1 arsec, but just with GALAXY or QSO classification in the DR16, is this OK?)}
%None of these sources were included in the Roma-BZCAT Multifrequency Catalogue of Blazars.

To find strong CL candidates, we performed a sky crossmatch within 1 arcsec between our type 2 sample and the sources reported by the ZTF alerts that were classified primarily as AGN or QSO according to the ALeRCE LCC, which led to 60 sources (see Table~\ref{tab: sample}).
Of these, $\sim$30 possible misclassifications arise from a visual examination of the optical light curves and the SDSS spectra, so the further cleaning of the sample will be addressed in a forthcoming paper. %On the one hand, some sources show obvious BELs in their optical spectra and were then misclassified in the catalogues. On the other hand, some could have been misclassified by the ALeRCE LCC due to a small number of data points or transient events in the reference image used to construct the difference images (mostly supernovae).
%is crucial to clean the final sample, since some of the selected candidates could have been miss-classified. On the one hand, some sources show obvious BELs in their optical spectra that were missed by the SDSS pipeline. %and 20 sources seem to have BELs that do not meet the BROADLINE criteria( lines detected at the 10-sigma level with sigmas > 200 km/sec at the 5-sigma level). 

%\citep{masci2019}

This selection method results in a sample of 60 CL candidates that show a type 1-like variability in the ZTF light curve, but that were previously classified as type 2 by the absence of significant BELs in their spectra. We note that we are searching for AGNs that changed from a type 2 to a type 1 classification according to their optical spectral lines only, so the selection is not biased towards obscured/unobscured AGNs. In order to test and refine this selection method, the next step is to confirm the sample spectroscopically by quantifying the change of the BELs.
\begin{table}
    \caption{Subsample of the 60 CL candidates selected through ALeRCE. Ndet: number of alerts reported by the ZTF. Prob.: probability of the AGN classification according to the LCC. The full list is available in the online version of the manuscript.}
    \label{tab: sample}

    \begin{tabular}{ccccccccc}
       \hline

ZTF ID & Ra(deg) & Dec(deg) & z & Ndet & Prob. \\
	\hline
ZTF18aaiescp & 207.212824 & 57.646811 & 0.13 & 14 & 0.81  \\

ZTF18aaihyhz & 216.756184 & 48.109091 & 0.04 & 36 &  0.63  \\

ZTF18aaiwdzt & 199.483614 & 49.258656 & 0.09 & 11 & 0.61  \\

ZTF18aaoudgg & 221.975974 & 28.556699 & 0.16 & 653  & 0.95  \\

ZTF18aaqftos & 180.955088 & 60.888197 & 0.07 & 345  & 0.41  \\

	    \hline
\end{tabular}

\end{table}

\section{Spectroscopic follow-up}
\subsection{Observations and data reduction}
With the aim of evaluating the selection method, we were allocated one night by the Chilean Time Allocation Committee (CNTAC) to observe a sample of five CL candidates with the Goodman High Throughput Spectrograph (GTHS) at the SOAR-4.1m telescope. The spectra were observed on November 2nd, 2021, using the red camera with the 400 lines/mm grating and the M2 filter (500–905 nm), a slit width of 1 arcsec, in normal readout and binning of 2x2. The total exposure time was 1h per source, divided in 3 different observations of 20 min each. On the other hand, we were also allocated observing time with FORS2-VLT at the UT1 Cassegrain focus to monitor a CL AGN event in real time. Since this AGN belongs to the same sample of candidates presented, we include here the first spectrum of the monitoring series, noting that the following spectra also contain significant BELs. The spectrum was taken on November 11th, 2021, with GRIS$\_$300V+10 (GC435 filter), 1 arcsec slit, MIT red-optimized CCD, 100kHz 2x2 high readout mode. We took two different exposures of 12.5 min each. For the SOAR/GTHS sources, we used the master bias and flats produced by the SOAR pipeline\footnote{\url{https://github.com/soar-telescope/goodman_pipeline}}, and the wavelength and flux calibration was obtained using the standard IRAF routines. In the case of the UT1/FORS2 object, we used the ESOReflex pipeline \citep{esoreflex}. Both sets of observations were corrected by telluric bands using the spectrum of the standard star and the telluric task from IRAF. The spectra were corrected by redshift using the information from the SDSS database. The details of the observations are reported in Table~\ref{tab: observations}.

\begin{table*}
	\centering
	    \begin{adjustbox}{max width=0.97\textwidth}
	    \begin{threeparttable}
	\caption{Sample of the CL candidates observed with \tnote{a}: SOAR/GTHS   and \tnote{b}:UT1/FORS2, and results of the spectral fitting. z: redshift from the SDSS database. EW: equivalent width of the emission lines. $\Delta_{t}$: time difference between the SDSS and the SOAR/GTHS or UT1/FORS2 spectra. \tnote{c}: derived properties for new (top) and archival (bottom) spectra. Slash (/): the broad component is not needed in the fit. \textbf{Bold}: sources with $>3\sigma$ change in the EW of H$_{\alpha}$ and H$_{\beta}$. Errors on derived quantities reflect the statistical uncertainties only, the associated systematic error is 0.5 dex on the mass and 0.3 dex on the bolometric luminosity used to calculate $\lambda_{\rm Edd}$.}
	\label{tab: observations}

	\begin{tabular}{ccccccccccccc} % number of columns, alignment for each
			\hline
		ZTF ID  & SDSS name  & z & epoch & SNR$_{5100}$\tnote{c} & EW$_{H_{\alpha}}$\tnote{c} & EW$_{H_{\beta}}$\tnote{c}   & H$_{\alpha}$/OIII\tnote{c}& $\Delta_{t}$&log M$_{\rm BH}$&$\lambda_{\rm Edd}$\\
		  &  &   & & & \AA &\AA   &&yrs& \(\textup{M}_\odot\) \\
		\hline
	\textbf{ZTF19abixawb}\tnote{a} & SDSS J001014.86+000820.7 & 0.1022 &new& 33 &47.0$^{+10}_{-0.3}$ &5.8$^{+5}_{-0.6}$&10.1$\pm$0.5&21.17&8.0 $\pm$0.2&0.013$\pm$0.007\\
		&&&old& 14 &8$^{+3}_{-1}$&/&2.3$\pm$0.5& \\
		\hline
		\textbf{ZTF20abshfkf}\tnote{a}& SDSS J011311.82+013542.4 &  0.2375 &new& 24 &133$\pm$2&29 $\pm$1 &5.8$\pm$0.4&6.13&8.0$\pm$0.1&0.03$\pm$0.01\\
		&&&&old& 18 &7$^{+20}_{-4}$&0$^{+2}$&0.2 $\pm$0.2& \\
		\hline
		\textbf{ZTF18accdhxv}\tnote{b}& SDSS J075544.35+192336.3 &0.1083 &new& 91 &90$^{+3}_{-20}$&15$^{+1}_{-10}$ &5.1$\pm$0.4& 17.02&8.5$\pm$0.1&0.005$\pm$0.001\\
		&&&old& 20 &21$^{+9}_{-1}$&1$\pm$1& 0.97$\pm$0.08 \\
		\hline
		\textbf{ZTF19aalxuyo}\tnote{a} & SDSS J081240.76+071528.5  & 0.0849 &new& 36 &54$^{+1}_{-3}$&4$\pm$1&9.8$\pm$0.3&17.64&7.76$\pm$0.04&0.027$\pm$0.005 \\
		&&&old& 27 &2.2$^{+5}_{-0.4}$&/&0.3$\pm$0.2\\
		\hline
		ZTF19aaxdiui\tnote{a} & SDSS J214046.03+091631.7 &  0.4030 &new & 17 &--&4$\pm$1&--&11.08&--&--\\ 
		&&&old& 17 &--&/&--& & \\
		\hline
		ZTF18abtizze\tnote{a} & SDSS J215055.73-010654.1 &  0.0879 &new& 24 &13$^{+1}_{-4}$ &/&0.42$\pm$0.05&17.39&7.4$\pm$0.2&0.005$\pm$0.003\\
		&&&old& 11 &3$^{+21}_{-1}$&/&0.1$\pm$0.1& \\
		\hline
		
	\end{tabular}
    
       \end{threeparttable}
       \end{adjustbox}

\end{table*}

\subsection{Spectral analysis}

\begin{figure*}
\centering
\begin{tabular}{ccc}

\includegraphics[width=0.38\textwidth]{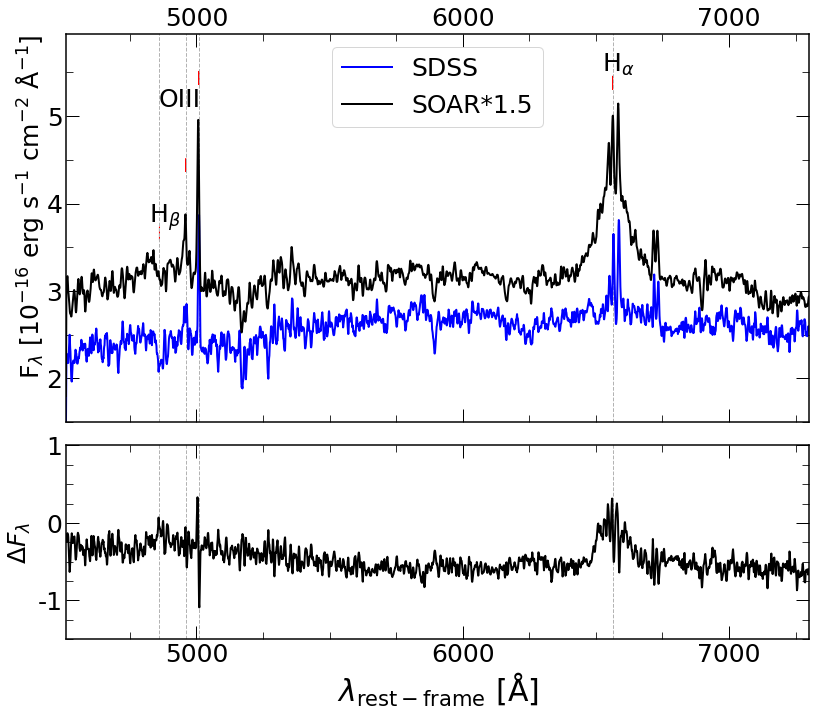} &
\includegraphics[width=0.38\textwidth]{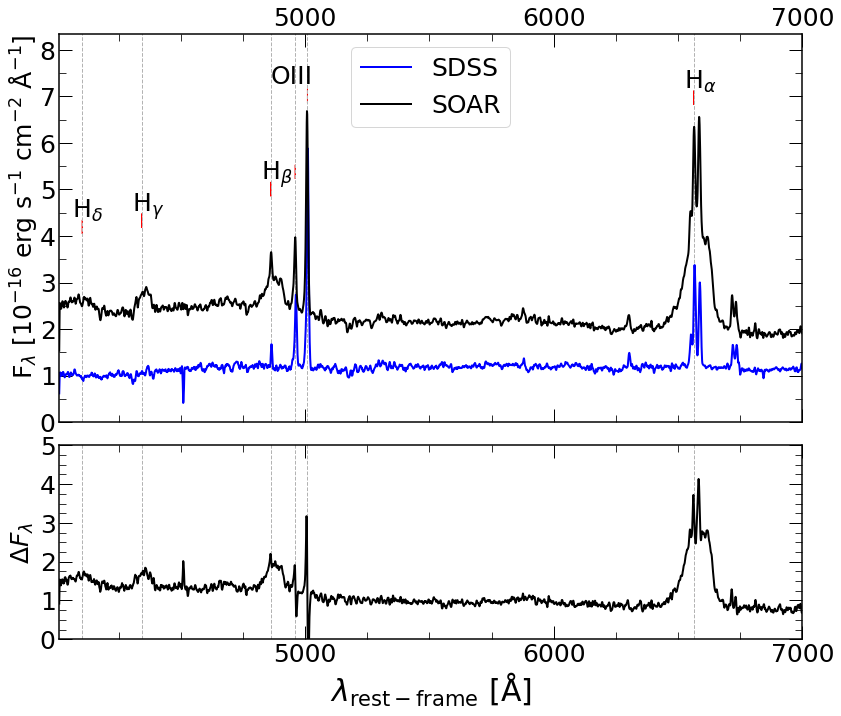}  \\
 \textbf{(a)ZTF19abixawb} & \textbf{(b)ZTF20abshfkf} 
\end{tabular}

\begin{tabular}{ccc}
\includegraphics[width=0.38\textwidth]{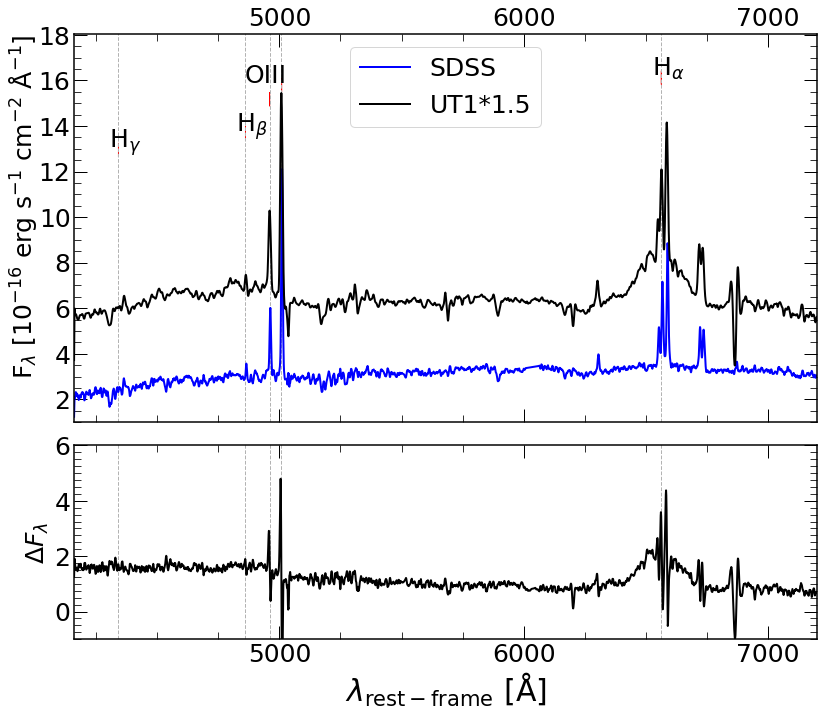} &
\includegraphics[width=0.38\textwidth]{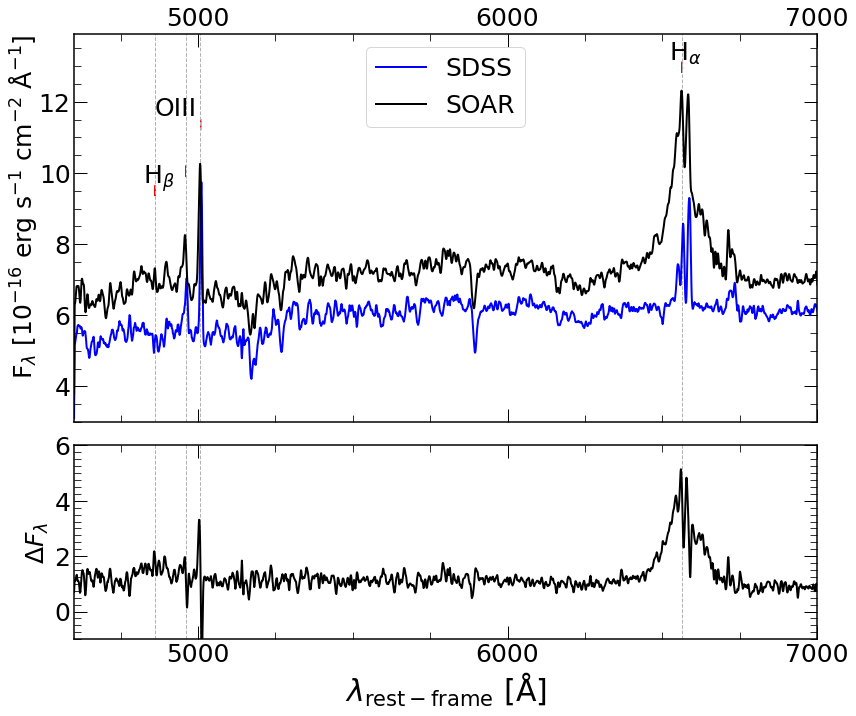}\\
   \textbf{(c)ZTF18accdhxv} &\textbf{(d)ZTF19aalxuyo} 
\end{tabular}

\begin{tabular}{ccc}
\includegraphics[width=0.38\textwidth]{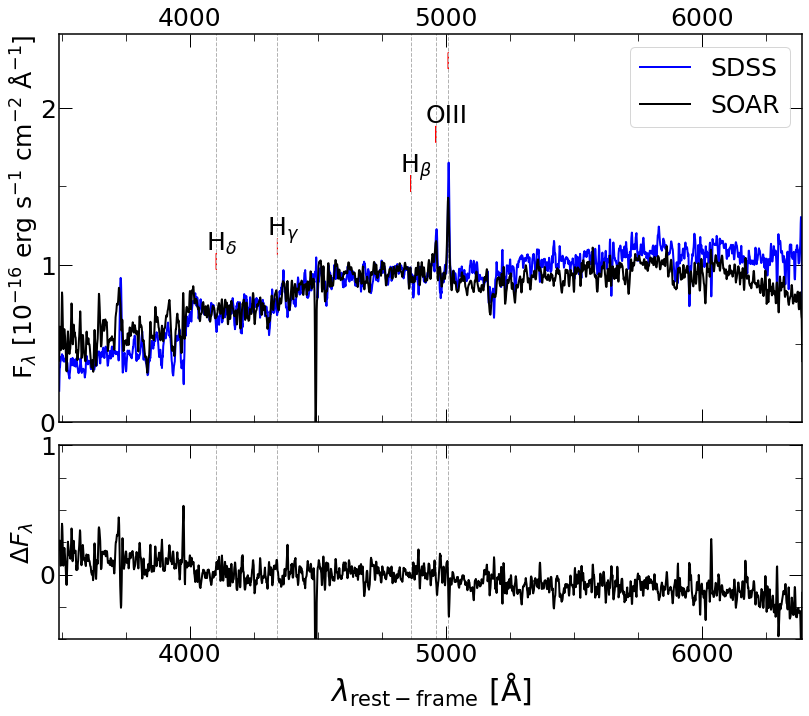} &
\includegraphics[width=0.38\textwidth]{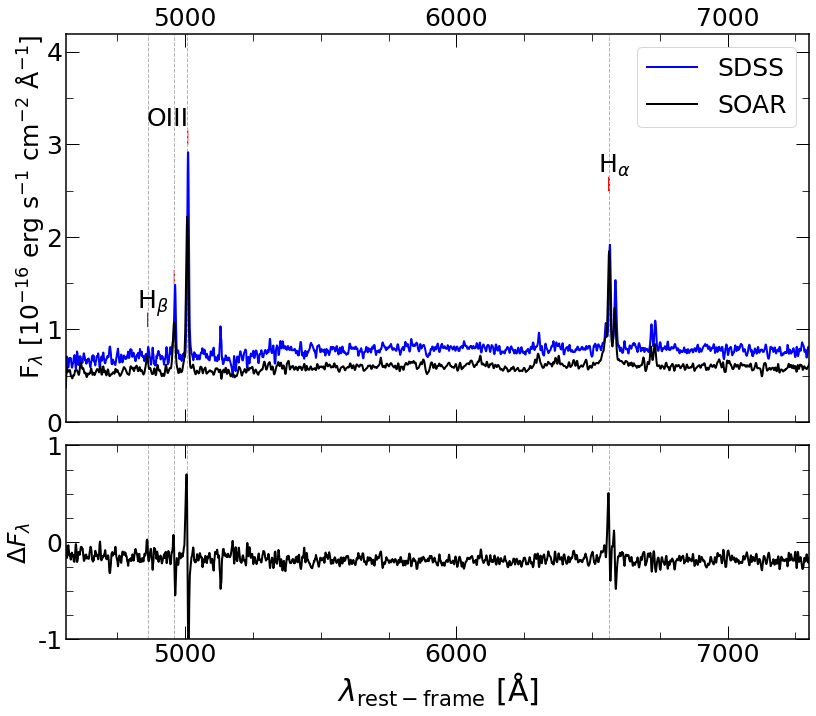}   \\
\textbf{(e)ZTF19aaxdiui}  & \textbf{(f)ZTF18abtizze}  
\end{tabular}

\caption{Archival (blue line) and new optical spectra (black line) for the subsample of CL candidates identified in this work, scaled to the flux of [OIII] $\lambda$=5007 \AA\ in the earliest spectra. The lower plots show the difference between the new and the old spectra. All spectra are smoothed with a 5 \AA\  box filter.}
\label{fig: spectra}
\end{figure*}

We analysed the new SOAR/GTHS and UT1/FORS2 observations and the archival SDSS spectra to compare the strength of the BELs in different epochs. In Figure~\ref{fig: spectra}  we present the plots of the average flux and difference spectra of the six CL candidates observed. It can be seen that ZTF19abixawb, ZTF20abshfkf, ZTF19aalxuyo and ZTF18accdhxv show a clear change both in the continuum and in the H$_{\alpha}$ and  H$_{\beta}$ BELs. To quantify the changes of the emission lines, we fitted the spectra using the Penalized Pixel-Fitting (pPXF) software \citep{cappellari2017improving}.  The SDSS spectra were fitted using the MILES library \citep{vazdekis2010evolutionary} for the stellar pseudocontinuum component, a set of power law models for the accretion disc contribution and two components for the emission lines, one with both permitted and forbidden lines to model the narrow emission and one just with the permitted lines to model the BELs. The second epoch spectra were fitted with the same components, using the stellar populations models from the best-fit to the SDSS spectra. To account for the fitting errors a total of 50 Monte Carlo simulations were performed for each spectrum using the residual of the best-fit to generate random noise. This noise was then added to the original spectrum and fitted with the same procedure. The errors reported for the EW in Table~\ref{tab: observations} correspond to the 10 and 90 percentiles of the simulations results. \\
For the sources with broad H$_\alpha$ emission, we estimated the black hole masses (M$_{\rm BH}$) and continuum luminosity at 5100 \AA~ ($L_{5100}$)  following the approach outlined in \citet{reines2013}, using the FWHM and luminosity of broad H$_\alpha$ obtained from the new spectra. With these values, we computed the current Eddington ratios $\lambda_{\rm   \rm Edd}=L_{\rm bol}/L_{\rm Edd}$, where $L_{\rm Edd} = 1.5 \times
10^{38}(M_{\rm BH} / M_\odot)$ erg s$^{-1}$ is the Eddington luminosity and  $L_{\rm bol}$ is the bolometric luminosity defined as $L_{\rm bol}=40 \cdot (L_{5100}/10^{42})^{-0.2}$ according to \citet{2019netzer}. %We note that, as in the SDSS spectra the signal to noise is lower and the BELs are very faint or nonexistent, the values for the Eddington ratios in the archival spectra were
The results of the spectral fitting and the derived physical quantities are shown in Table~\ref{tab: observations}.

\section{Discussion}
The selection of our CL candidates sample is based on current type 1 optical variability reported by the ZTF and previous type 2 spectral classification. Thus, this method leads to sources that showed very weak or absent BELs and now behave as normal type 1 AGNs, with obvious BELs in their optical spectra. Our sample is similar to the CL AGNs identified by comparing repeated spectroscopy as in \citet{yang2018}, and differs from the changing-state AGNs in that the latter usually refer to changes in the flux of existent BELs. 
Our method results in 60 CL candidates, from which half appear to be misclassified based on visual inspection of their spectra and light curves, leading to a list of $\sim$ 30 promising CL candidates. Here we confirm a significant spectral change (a $>3 \sigma$ change in the EW of H$_{\alpha}$ and H$_{\beta}$) in four out of the six sources that we re-observed, which were chosen from our list of promising candidates based on observability considerations (i.e. that were observable on the same night we had been allocated). Despite the small size of this sub-sample, our first results suggest a success rate (SR) that is comparable to the 70 per cent SR reported in \citet{2021hon}, which is the most successful search method for CL AGNs to date in comparison to previous works with variability selection criteria \citep[15-35 per cent, e.g.][]{yang2018,macleod2019}.

With the results of this work, we can also make a lower limit estimate of the frequency of type 2 to type 1 transitions. There are 29943 type 2 sources (\textsc{Broad$\_$Type} N or K in MILLIQUAS) that can be detected by the ZTF alert stream (dec$>$-28 deg and r $<$ 21 mag). From these, 158 are classified as QSO or AGN by the LCC. As pointed out in Section \ref{sec:sample}, we need to take into consideration misclasifications in the parent sample, which could reach  30-50 per cent. This high probability comes from the fact that we are searching for oddities, so the frequency of misclassifications, which is overall small, matches the frequency of real candidates, which is also very small. Considering 50 per cent misclassifications and a minimum success rate of 2/3 (as found in this work) for spectra separated by an average of 15 years, we estimate a fraction of 0.18 per cent of type 2 AGN turning into type 1 per 15 years, or 0.01 per cent per year, that could be detected with this method. This number must be corrected by the percentage of sources that actually are variable enough to generate ZTF alerts (10 per cent for known type 1 AGNs), leading to a lower estimate of 0.12 per cent type 2 sources changing into type 1 per year (1.8 per cent in 15 years). This lower limit is similar to the value reported by \citet{2021hon}, who found a minimum turn-on CL AGN rate of 3 per cent over $\sim$15 yr.

%Considering a success rate of 2/3 (as found in this work) for spectra separated by an average of 15 years, we estimate a fraction of $0.35\%$ of type 2 AGN turning into type 1 per 15 years, or $0.02\%$ per year, that could be detected with this method. This number must be corrected by the percentage of sources that actually are variable enough to generate alerts (10$\%$ for known type 1 AGNs), leading to an upper estimate of $0.2\%$ type 2 sources changing into type 1 per year. We note that to get a more plausible estimate, we need to take into consideration missclasifications in the parent sample, which could fluctuate within the range of 30-50$\%$. This high probability comes from the fact that we are searching for oddities, so the frequency of missclassifications, which is overall small, matches the frequency of real candidates, which is also very small.

Our results indicate that the sources in this work are currently accreting at a few per cent L$_{\rm Edd}$, which is close to the value for the state transitions observed in XRB, and suggests an accretion state change as the origin of the CL phenomena in these sources \citep{Noda2018, ross2018, hutsemekers2019, graham2020, 2021guolo}. In these transitions, the soft/hard accretion states are caused by pronounced changes in the accretion disc contribution to the total radiative output. According to the latest AGN accretion-oriented diagrams, the soft state in AGNs would include broad-line Seyferts, showing highly excited gas and radio-quiet cores consistent with disc-dominated nuclei, while most true Seyfert 2 nuclei and the bright LINERs would show low excitation at high accretion luminosities and could be identified with the bright hard and intermediate states \citep{fernandez_ontiveros2021}. Here, our preliminary fits to the optical spectra suggest a significant increase of 40-70 per cent in the AGN component for ZTF19abixawb and ZTF20abshfkf, in agreement with the scenario where our sources transitioned from true type 2 AGNs, with some or negligible contribution from a cold accretion disc, %and a radiatively inefficient accretion flow or RIAF, 
to type 1 disc dominated sources. Incidentally, the light curves of these sources show persistent stochastic variations as typical type 1 AGNs, whereas the other two confirmed CLs, ZTF19aalxuyo and ZTF18accdhxv, show an increasing optical flux consistent with a transition to a disc dominated state. Independently of the AGN continuum contribution (that can be very uncertain), the  H$_\alpha/$OIII ratios of the four CL AGNs are now between 5 and 30 times larger than in the archival spectra (see Table~\ref{tab: observations}), which most likely reveals the change in the ionising flux. 
On the contrary, the two candidates that were not confirmed as CL show a strong declining trend over the last 2-3 years before the new spectra were taken. This suggests that we could have missed the transitions from true type 2 to type 1 and again to type 2. In fact, preliminary fits to the light curve of ZTF18abtizze point to a TDE as the origin of the optical variations, but further analysis is required to better understand the nature of this source. We note that for the other not-confirmed CL AGN, ZTF19aaxdiui, H$_\alpha$ falls out of the observed wavelength range, but the EW$_{H_{\beta}}$ is the same as for the CL ZTF19aalxuyo (which displays very significant changes in H$_\alpha$). Therefore,  we cannot rule out this source is actually a CL AGN. %and since the changes in H$_\beta$ are more difficult to detect,
\section{Summary and conclusions}

We have selected a sample of 60 CL AGNs candidates that were classified spectroscopically as type 2, but currently show significant photometric variations according to the ALeRCE LCC. To test our selection criteria, we re-observed six of these sources with SOAR and VLT telescopes. By comparing these and archival SDSS observations we find that at least four of the sources have experienced a significant ($>3\sigma$) increase in the emission of their Balmer BELs, which implies a promising success rate of $\geq66$ per cent. Our spectral fits suggest the CL AGNs are currently accreting at a few per cent L$_{\rm Edd}$, which is consistent with an accretion state change as the origin of the CL phenomena in these sources. To the best of our knowledge, this is the first time the ALeRCE broker LCC has been used to select new potential CL AGNs candidates, and we are currently re-observing the full promising CL sample to refine the statistics on the frequency of type 2 to type 1 transitions, which will be presented in a forthcoming paper. In future work, we will improve the completeness of the candidate list by characterising light curves that use all available epochs of ZTF instead of only using the ZTF alert stream. This will include more of the lower-variability AGNs, which comprise 90 per cent of known type 1 sources.

\section*{Acknowledgements}
We acknowledge financial support from ANID Becas 21200718 and 21212344 (ELN, SB), Millenium Nucleus NCN$19\_058$ TITANS (PA, MLMA), ICN12$\_$009 MAS (LHG), FONDECYT Nº 3200250 and 1201748 (PSS, PL). ELN acknowledges Caltech for its hospitality. Based on observations collected at the European Southern Observatory under ESO programme 108.22BA.001, at the Southern Astrophysical Research telescope under CNTAC programme CN2021B-16, and with the Samuel Oschin Telescope 48-inch and the 60-inch Telescope at the Palomar Observatory as part of the ZTF project.
%which is a joint project of the Minist\'{e}rio da Ci\^{e}ncia, Tecnologia e Inova\c{c}\~{o}es (MCTI/LNA) do Brasil, the US National Science Foundation’s NOIRLab, the University of North Carolina at Chapel Hill (UNC), and Michigan State University (MSU).
%This research made use of Astropy, a community-developed core Python package for Astronomy (Astropy Collaboration, 2013, 2018).

%%%%%%%%%%%%%%%%%%%%%%%%%%%%%%%%%%%%%%%%%%%%%%%%%%
\section*{Data Availability}

The SDSS data underlying this article were accessed from SDSS DR16 (\url{http://skyserver.sdss.org/dr16}). The UT1/FORS2 and SOAR/GTHS data underlying this article will be shared on reasonable request to the corresponding author.% data underlying this article are subject to an embargo of 12 months since the observations were taken. Once it expires, the data will be available in the ESO archive  (\url{http://archive.eso.org/eso/eso_archive_main.html})



\bibliographystyle{mnras}
\bibliography{example} % if your bibtex file is called example.bib

%%%%%%%%%%%%%%%%%%%%%%%%%%%%%%%%%%%%%%%%%%%%%%%%%%

% Don't change these lines
\bsp	% typesetting comment
\label{lastpage}
\end{document}